\def\year{2021}\relax
\documentclass[letterpaper]{article} 
\usepackage{aaai21}  
\usepackage{times}  
\usepackage{helvet} 
\usepackage{courier}  
\usepackage[hyphens]{url}  
\usepackage{multirow}
\usepackage{xcolor}
\usepackage{booktabs}
\usepackage{tabularx}
\usepackage{bm}
\usepackage{balance}
\usepackage{multirow}

\usepackage{float}

\usepackage{graphicx}               
\usepackage{natbib}                 
\usepackage{caption}                
\title{An In-depth Review  of Privacy Concerns Raised by  the COVID-19 Pandemic}

\author {
    Jiaqi Wang\\
}

\affiliations{
    Penn State University \\
    jqwang@psu.edu
}

\begin{document}

\maketitle
\begin{abstract}
COVID-19 has hugely changed our lives, work, and interactions with people. With more and more online activities, people are easily exposed to privacy threats. In this paper, we explore how users self-disclose on social media and privacy concerns raised from these behaviors. Based on recent news, techniques, and research, we indicate three   increasing privacy threats caused by the COVID-19 pandemic. After that, we provide a systematic analysis of potential privacy issues  related to the COVID pandemic. Furthermore, we propose a series of research directions about online user self-disclosure and privacy issues for future work as well as possible solutions.
\end{abstract}
\section{Introduction}
COVID-19 has spread across the world and affected how people work, live, and interact with each other. People are recommended or required to work remotely, quarantine at home, and keep social distance. Under these circumstances, people expect more interactions with others via social media platforms, which has led to  a huge increase of social media usage \cite{holmes_2020}. Based on a study \cite{kantar} of 25,000 consumers across 30 markets published on April 3rd, 2020, WhatsApp has seen a 40\% increase in usage; in the early phase of the pandemic usage increases 27\%, in mid-phase 41\% and countries in the late phase of the pandemic see an increase of 51\%; Facebook usage has increased 37\%. China experienced a 58\% increase in usage of local social media apps including Wechat and Weibo. Another study of 4500 \textit{Influenster} community members, most of respondents agreed that their social media consumption (72\%) and posting (43\%) have increased during the pandemic. Moreover, TikTok, one of new social media platforms, was used by the largest share of 
teenagers (48\%), overtaking even Instagram (47\%) from March, 2020 to April, 2020 \cite{s_2020}.

One  possible reason is that people are searching for alternative approaches to interact with others to stay mentally healthy. People generate content, comment content, forward content, and communicate with others on social media platforms. To increase  a sense of  intimacy with others, people share details of their lives with text, pictures, videos, live video streaming, etc. To a great extent, the content can reveal personal private information including age, gender, location, race, etc. Compared with interactions in the real world, self-disclosure information can more easily be propagated, searched, saved, and 
even processed on social media. The increasing and more abundant self-disclosure may cause unpredictable and unacceptable privacy disclosure to users online. Furthermore, a recent research shows that people's mental health problems are prevalent because of social media exposure \cite{gao2020mental} itself, which means the expected results might be on the contrary to the mental health cure.

However, the pandemic is changing people's sensitivity and attitude to privacy including what and how personal information can be disclosed \cite{nabity2020inside}. Discussion about COVID-19 may include basic personal information, travel schedule, test results, symptom description, and medicine in use. These acts of  self-disclosure reveal a lot of sensitive information that people are not willing to share previously \cite{kordzadeh2017communicating}. For example, health status and detailed description of individual body information are shared to ask for comparison, suggestions or pre-diagnosis. Some communities even encourage people to share more personal information related to COVID-19 in the name of society responsibility without clarifying the boundary of gathered information and how to use the collected data. Based on the observation, users would sacrifice personal information to a unprecedented degree to help the society back to the expected normal status. Recent work \cite{blose2020privacy} provides early evidence that the situational factors caused by COVID-19 may affect people's self-disclosures and privacy calculus. 
\begin{figure*}[htbp]
\centering
\includegraphics[width=0.75\linewidth]{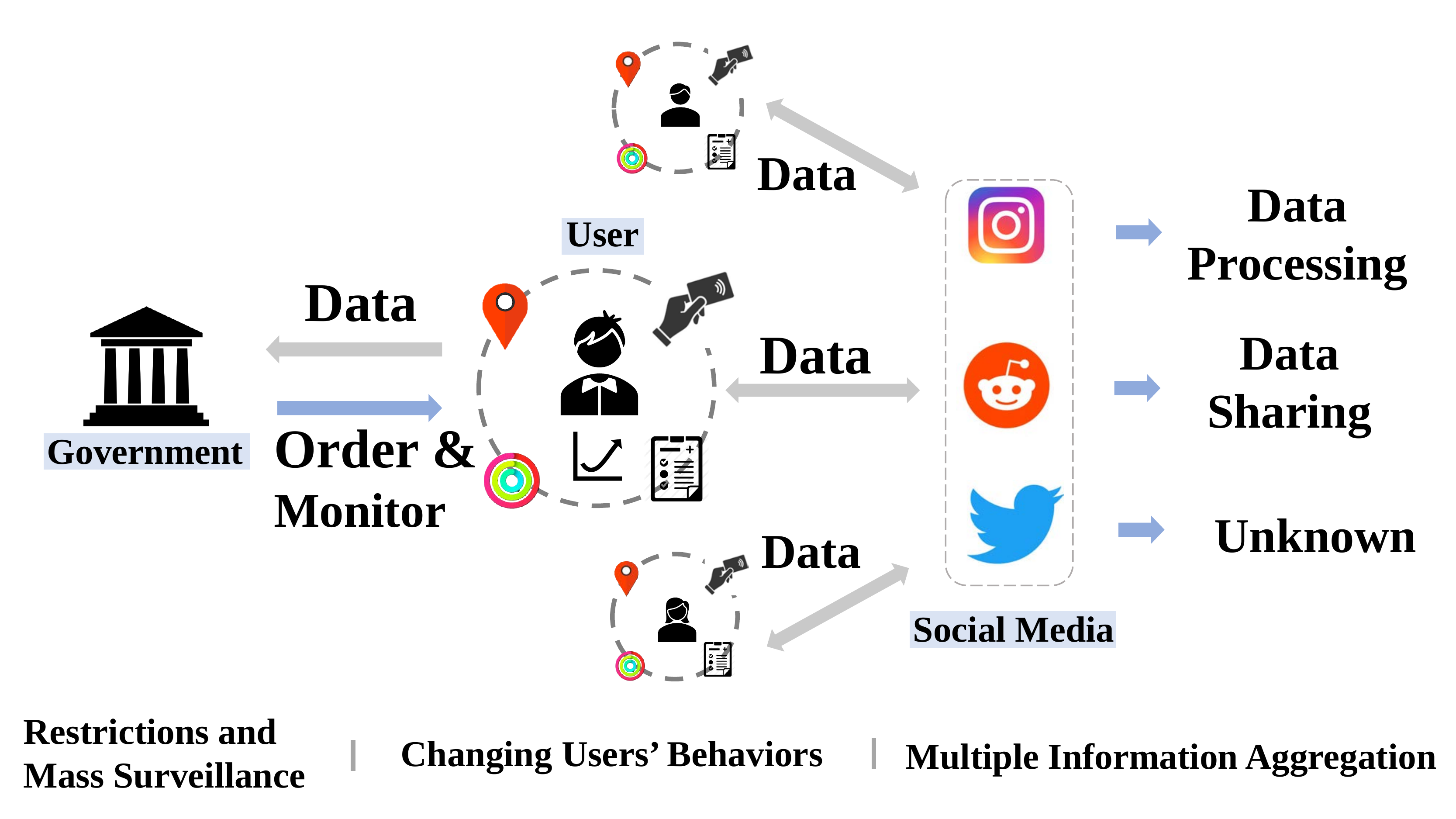}
\caption{A Systematic Overview of Privacy Threats from Multiple Domains Related to the COVID-19 Pandemic}
\label{fig:system}
\end{figure*}
There is another  issue we need to pay attention to. Along with the COVID-19 pandemic, 2020 the United States presidential elections started from February and ends in November. Noting that the date when United States officially declared the COVID-19 pandemic as a national emergency is March 13 and the first statewide "stay-at-home" order was issued at California is March 16. That time is approximately only one month later than the early voting in February. During the whole process of the presidential election, people are isolated at home and keep social distance in essential activities at most time.  People have participated extensively in  political discussions, and actively engaged in social media pushed by a highly divisive environment.  This is likely linked to users disclosing sensitive information including but not limited to political stand, home address, and family relative information. The potential privacy harms to users in the context of political debates have been studied before \cite{rubinstein2014voter}. However,  this election has introduced even additional  situational factors, as it happened in the middle of a pandemic.
 
Information sources across multiple social media may cause serious user privacy issues and unclear self-disclosures under the chaotic interactions with natural and social environment.  Advanced machine learning and data mining techniques investigate non-obvious  relationships and search hidden data patterns, which can provide insights to the data owners and external parties for unknown analysis \cite{chamikara2020efficient}.

In the following, we first summarize and analyze emerging  privacy threats triggered by or enhanced by the COVID-19 Pandemic. Based on our findings, we provide a high-level comprehensive analysis of privacy from multiple domains, propose related potential research directions,and conclude implications for future online public privacy in crisis.. Finally, we discuss possible solutions of proposed research questions.

\section{Increasing Privacy Threats due to the COVID-19 Pandemic}
\subsection{Mass Surveillance}
 There is an ongoing public conversation   about whether and under what circumstances the United States should embrace a surveillance program for COVID-19 \cite{ram2020mass}. Here, we focus  on what tools the government and companies are leveraging from the phenomenon perspective. There is increasing surveillance over people's daily behaviors from the government and companies during the COVID-19 pandemic in the name of monitoring and tracing the virus spread \cite{hussein2020digital}. Many countries and companies are leveraging people's personal data (location, body temperature, facial information, etc.), which is collected by cell phones, traffic cameras, and other sensors, to track human mobility, identify individuals with risk, and monitor the disease spread \cite{singer_sang-hun_2020}. In the United Kingdom and India, smart city infrastructure has been re-used to monitor the people's social distance. In China, people can download a cell phone application that can tell whether they have been exposed to COVID-19 by analyzing the collected location data and local infection situation \cite{bbc_news_2020}. In the United States, Apple and Google provided a contact tracing application for their mobile users as well with bluetooth specification \cite{apple_google_1} and cryptography specification \cite{apple_google_2}. However, as a key part of the extension of the surveillance state, researchers stated that the anonymized data is not always anonymous and location data can exacerbate inequality. \cite{frith2020covid19}.
\subsection{Data Usage across Multiple Platforms}
During the COVID-19 pandemic, people spent extensive  time online communicating, generating content, and engaging in other activities. With the development of data science techniques, people have more computational power and various channels to collect, process, and share data. There have already a lot of released open datasets focusing on different aspects related to the COVID-19 \cite{blose2020privacy, chen2020covid, pepe2020covid,cohen2020covid,cheng2020covid,dong2020interactive}. Many social media platforms provide APIs for people to acquire data, such as Twitter \footnote{https://developer.twitter.com/en/docs/twitter-api} and Reddit \footnote{https://www.reddit.com/dev/api/}. Those APIs lower the barrier to access social media data. However, we can not fully prevent  malicious usage of the collected data. 

At the same time, more digital records and accounts containing sensitive information are being created online, for example, online shopping accounts \cite{brough2020consumer} and other services that are brought online.  Online users may not be fully aware of the fact their private information can be collected, shared, and used in an unexpected way \cite{malandrino2013privacy}.   Many users may have more than one accounts on social media. How to measure privacy disclosure score based on the information across multiple social networks has been  discussed \cite{aghasian2017scoring} extensively. \citeauthor{zola2019social} explored a cross-source cross-domain sentiment analysis with training data from Amazon and Tripadvisor and testing on the data from Facebook and Twitter \cite{zola2019social}. 
\subsection{Change of Individual Privacy Calculus}
Another observed phenomenon and potential concern  is the change of individuals' perception  to self-disclosure and privacy. Individual-level behavior during the pandemic is a result of voluntary and government-enforced behavioral change \cite{farooq2020impact}. From the individual perspective, people are calibrating  their behavior  between  information acquisition and privacy loss. Users may have different attitudes and sensitivity to their privacy and self-disclosure during the pandemic \cite{fahey2020covid}. People would more easily sacrifice their private health status information to get suggestions, pre-diagnosis, or contribute to what the government appeals during the COVID-19 pandemic, especially in Asia \cite{cha2020asia}. Discussing personal health status, symptom, and test results on social media has become more common. Governments and companies provide convenient tools for people to update their personal information and implicitly convince people that the behaviors are a contribution to the public good \cite{nabity2020inside}. However, to my best knowledge, there are not enough official files to remind people about individual privacy issues or broadcast basic knowledge of data usage for people during the COVID pandemic. 

A systematic overview of privacy issues from different aspects during the COVID-19 Pandemic is shown in Figure \ref{fig:system}.

\section{Post-pandemic Potential Privacy Risks }

\subsection{Over-collected Data Abuse}
The COVID-19 pandemic  has promoted the development of e-commerce, online education, social media platforms, smart phone applications, and related virtual service. Due to the health emergency, many countries relax the regulation restrictions or cooperate with companies to put the public security in the first place by collecting and analyzing data to support governmental prevention decision making. The governments could leverage  contact tracing information to monitor and analyze citizens' behaviors, e.g.    LGBT people identification in South Korea \cite{fahey2020covid}. Some countries will put pressure on their companies to release
the collected data and provide data analysis on the involved users. The European Commission has invited telecommunications companies to make their metadata available \cite{turner_2020}.

\begin{figure}[htbp]
\centering
\includegraphics[width=0.9\linewidth]{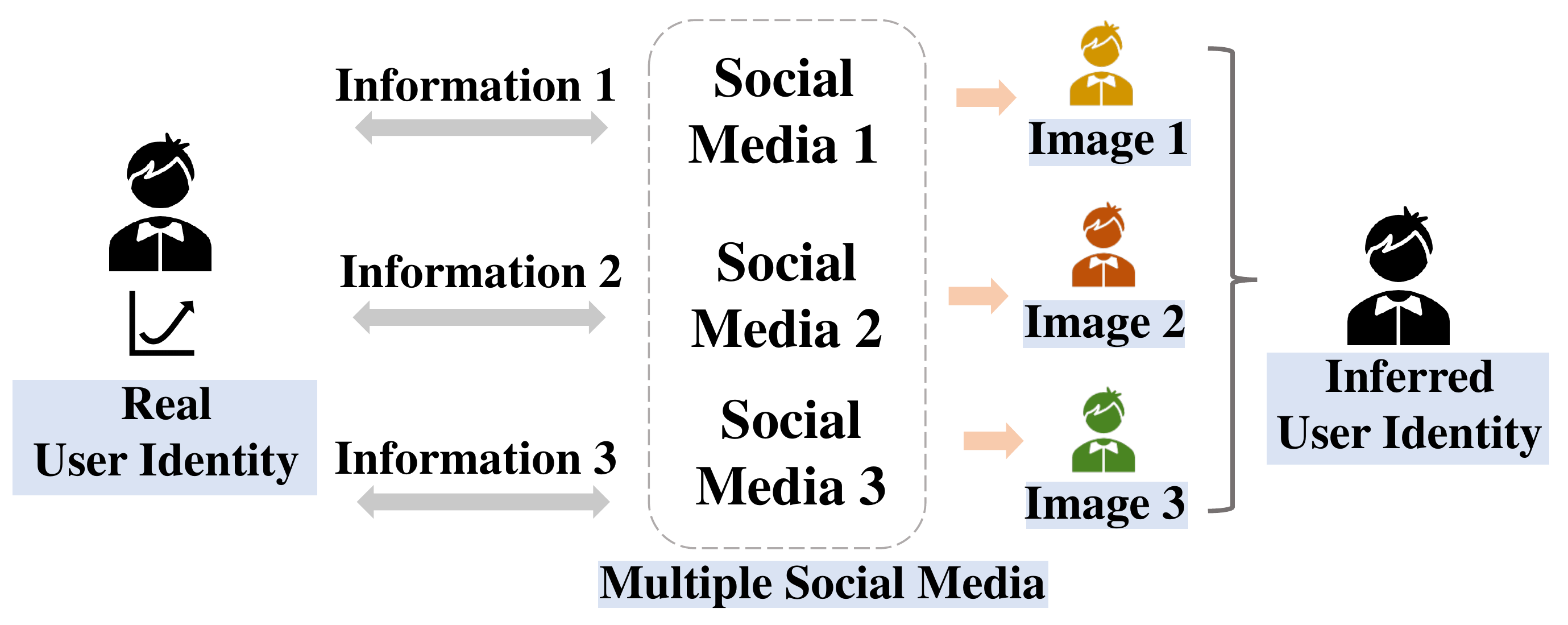}
\caption{Users Potential Privacy Risks: User Identity Inference based on Multiple Social Media. For each social media, one user would self-disclose part of personal information, for example, Information 1, Information 2, and Information 3. According to the disclosed information, one user can be treated as fuzzy image with released and limited inferred information on one social media, for example, Image 1, Image 2, and Image 3. However, given multiple social media data of one user and advanced across-platform data processing techniques, data can be aggregated to infer a more accurate user  identity with detailed personal information.}
\label{fig:infer}
\end{figure}

Tech companies, including Instagram, Twitter, Facebook, and etc., can abuse this detailed  data sets of individually, by selling, processing it to derive sensitive information, or sharing it inappropriately. Relying on powerful computational resources such as GPU clusters, a huge amount of data, and advanced data processing techniques, users behaviors can be described, modelled, and predicted accurately without any consideration for users' privacy. For example, an example of user behavior identification and prediction across multiple social media platforms is shown in Figure \ref{fig:infer}. Moreover, people share content via text, pictures, video, live streaming, and other formats, which can provide comprehensive information of users. Online interactions, e.g., ``Follow'', ``Hashtag'', ``Mention'', ``Reply'', can even reveal users' friends and relatives and create their social network structure. That would cause other related users' the privacy loss and over-disclosure and the propagation of the threat across the whole social media.

\begin{table*}[t]
\caption{Possible Research Directions and Questions about Privacy Issues and Self-disclosure related t Crisis On Social Media} \label{pval} 
\centering
\scalebox{1}{
\begin{tabularx}{\textwidth}{cX}
\toprule
    Research Directions       &      Research Questions    \\
          \midrule
\multirow{5}{*}{Self-disclosure Interaction and Propagation} & $\bullet$ How and to what extent users' self-disclosure behaviors can affect other related users on social media? \\
            & $\bullet$ How the self-disclosure behaviors propagate on the social media? \\
            & $\bullet$ To what extent the crisis would affect the user self-disclosure behaviors?\\
            & $\bullet$ How to find the balance point between the privacy preserving and self-disclosure to get enough and appropriate information in crisis?\\
            & $\bullet$ How to quantify   self-disclosure across multiple social media and provide a varying evaluation considering situational factors?\\
\midrule
\multirow{4}{*}{Public Privacy Concern and Attitude Tracing} &  $\bullet$ How to trace the public privacy attitude change to their current status?  \\
            & $\bullet$ How to design an appropriate data-driven mechanism and regulation to gather appropriate data and decrease the public privacy concern? \\
                        & $\bullet$ How to model the complex and dynamic observations considering users' privacy concern, users' behaviors, and the pandemic crisis?       \\
\midrule
\multirow{5}{*}{Mental Health in the COVID-19 Pandemic} & $\bullet$ How to find a balance   between keeping mental health and privacy  during the pandemic? \\
            & $\bullet$ How the mental health status, self-disclosure, and privacy concern affect each other? Certain self-disclosure can help users keep a good mental health, while it takes private concerns to users as well. \\
            & $\bullet$ During the health emergency crisis, considering users with different physical health status, would there be any differences of their mental health and online behaviors?\\
\midrule   
\multirow{5}{*}{Prevention, Prediction, and Protection} & $\bullet$ How to design a comprehensive mechanism to prevent over self-disclosure and privacy-disclosure according to complicated scenarios in crisis?\\
            & $\bullet$ How to predict  public behavior  and provide appropriate suggestions with limited access of data during the pandemic? \\
            & $\bullet$ How to protect users' provided data, protect the stability on social media, and establish social trust?\\
\midrule 
 \bottomrule                                   
\end{tabularx}}
\label{table:research}
\end{table*}

\subsection{Public Privacy Concern and Social Trustworthiness}
As the COVID-19 pandemic carries on,   debates and  laws   surrounding  surveillance capabilities  are at the forefront of many minds    \cite{ross_2020}. However, a majority of Americans said that they were concerned about how their personal data would be used by data collectors and they knew extremely little about the laws or regulations to protect their data privacy\cite{auxier_2020}. Many governments gather or even over-collect people's data during the pandemic via different approaches. There is a great possibility that they will not delete the collected personal data or even continue collecting the data without informing users. Another survey result in \cite{auxier_2020} shows that 69\%  U.S. adults thought they should have the right to have the medical data permanently deleted after necessary and legal usage. While people enjoy the benefit of pandemic tracking and controlling via the data-driven approach, it also raises public concerns for their individual privacy. \citeauthor{kye2020social} argued that the government actions do have a huge impact on social trust and government Trustworthiness. The temporal over-disclosed data and privacy data disclosure is gradually causing a stronger public privacy concern and challenging the government social trust.

\section{Potential Research about Pandemic-related Privacy Issues on Social Media}
Based on previous work and our discussion, we propose a set of related research directions (shown in Table \ref{table:research}) to understand  and explore further privacy issues at time of COVID. They include: (i) self-disclosure interaction and propagation; (ii) public privacy concern and attitude tracing; (iii) mental health; (iv) prevention, prediction, and protection in the COVID pandemic. For each research direction, we provide several related specific research questions in the table \ref{table:research} as well for future exploration.

\section{Conclusion}
The COVID-19 pandemic  has generated a lot of practical problems and research questions related to privacy issues in online settings. In this paper, we describe how the COVID-19 affects user behaviors on social media. After that, we discuss three increasing privacy threats due to the pandemic including mass surveillance, data usage across multiple platforms, and change of people's privacy calculus. Furthermore, we introduce possible privacy risk after the pandemic. Finally, we propose a set of related research topics for further study. There could be several possible  research directions: (i) appropriate and adaptive approaches to quantify self-disclosure and privacy combining peoples' comprehensive behaviors in multiple scenarios; (ii) mathematical and statistical models of privacy and human behaviors rather that can complement  data-driven approaches ; (iii) study the interactions between people's awareness and sensitivity of privacy and self-disclosure considering the changes of environment. Different people may have different initial attitudes towards their personal information and decide how much information they feel comfortable to self-disclose. The exploration of the hidden relation between privacy attitudes, self-disclosure behaviors, and the reaction got from the environment can help us understand humans' privacy-related behaviors better and provide comprehensive suggestions for privacy-preserving mechanism design.

\bibliography{cite}

\end{document}